\documentclass[aps,pre,reprint,superscriptaddress]{revtex4-1}

\usepackage[dvips]{graphicx}
\usepackage{bm,color}

\begin{document}


\title{Ultrafast Wave-Particle Energy Transfer in the Collapse of
  Standing Whistler Waves}


\author{Takayoshi Sano}
\email{sano@ile.osaka-u.ac.jp}
\affiliation{Institute of Laser Engineering, Osaka University, Suita,
  Osaka 565-0871, Japan} 

\author{Masayasu Hata}
\affiliation{Institute of Laser Engineering, Osaka University, Suita,
  Osaka 565-0871, Japan} 

\author{Daiki Kawahito}
\affiliation{Center for Energy Research, University of California San
  Diego, La Jolla, CA 92093-0417, USA}

\author{Kunioki Mima}
\affiliation{Institute of Laser Engineering, Osaka University, Suita,
  Osaka 565-0871, Japan} 
\affiliation{The Graduate School for the Creation of New Photonics
  Industries, Hamamatsu, Shizuoka 431-1202, Japan} 

\author{Yasuhiko Sentoku}
\affiliation{Institute of Laser Engineering, Osaka University, Suita,
  Osaka 565-0871, Japan} 


\date{November 13, 2019; accepted for publication in Physical Review E}

\begin{abstract}
{
Efficient energy transfer from electromagnetic waves to ions has been
demanded to control laboratory plasmas for various applications
and could be useful to understand the nature of space and
astrophysical plasmas.}
However, there exists a severe unsolved problem that most of the wave
energy is converted quickly to electrons, but not to ions.
Here, an energy conversion process to ions in overdense plasmas
associated with whistler waves is investigated by numerical
simulations and theoretical model.
Whistler waves propagating along a magnetic field in space and
laboratories often form the standing waves by the collision of
counter-propagating waves or through the reflection.    
We find that ions in the standing whistler waves acquire a large
amount of energy directly from the waves in a short timescale
comparable to the wave oscillation period. 
Thermalized ion temperature increases in proportion to the square of
the wave amplitude and becomes much higher than the electron
temperature in a wide range of wave-plasma conditions. 
This efficient ion-heating mechanism applies to various plasma
phenomena in space physics and fusion energy sciences. 
\end{abstract}


\maketitle


\section{Introduction}

Plasma acceleration and heating by electromagnetic waves is of great
importance in many research topics such as parametric instabilities
\cite{kruer03}, collisionless shocks and
turbulence \cite{lembege04}, planetary magnetospheres
\cite{stenzel99,stenzel16}, and inertial and magnetic confinement
fusion (ICF and MCF) \cite{lindl95,betti16,stacey05}.  
Among different types of waves, the whistler wave, which is a
low-frequency electromagnetic wave traveling along an external
magnetic field $B_{\rm ext}$, often plays a major role in the
generation of energetic particles \cite{stenzel99,stenzel16}.   
{
Whistler-mode chorus waves are one of the most intense plasma waves
observed in planetary magnetospheres
\cite{helliwell65,tsurutani74,coroniti80,hospodarsky08,artemyev16}
and expected as a promising mechanism to produce
relativistic electrons
\cite{horne05,omura07,summers07,thorne13}. 
Whistler waves are considered useful for inducing plasma currents
and heating electrons in tokamak devices for MCF
\cite{stacey05,prater14}
and also generated in laser plasmas of ICF experiments
\cite{mourenas98,taguchi17}.} 

The whistler wave is a right-hand circularly polarized (CP) light
permitted to exist when the electron cyclotron frequency $\omega_{ce}
= e B_{\rm ext} / m_e$ exceeds that of the electromagnetic wave
$\omega_0$, where $e$ is the elementary charge and $m_e$ is the
electron mass. 
The critical field strength $B_c$ is defined by $B_c \equiv m_e
\omega_0 / e$ assuming $\omega_0 = \omega_{ce}$. 
Note that the required magnetic field becomes weaker if the whistler
frequency is lower or the wavelength is longer. 

The whistler wave has interesting characteristics that give an
advantage to plasma heating processes.  
The most important feature is no cutoff density for the whistler
waves. 
{
Whistler waves can propagate inside of any density plasmas unless they
encounter a strong density gradient so that they interact 
directly even with overdense plasmas \cite{yang15,ma16,weng17}.}
Another critical fact is that a large electromotive potential, or
an electrostatic potential, in the
longitudinal direction appears in the standing wave of whistler-mode.
The standing waves are naturally excited by overlapping two
counter-propagating waves or by the reflection at the discontinuity of 
plasma density.
The rapid build-up of the electrostatic potential accelerates ions.
The amplitude of the potential energy is roughly given by $\psi \sim e v_w
B_w \lambda_w / (2 \pi)$, where $v_w$, $B_w$, and $\lambda_w$ are the amplitude of velocity, magnetic field, and wavelength of a whistler
wave.
The potential energy could be of the order of MeV for the relativistic
whistler cases, and thus it is an attractive source for the 
energy transfer from the waves to plasmas.
Nevertheless, the details of plasma acceleration and heating during
the interaction between the standing whistler waves and overdense
plasmas have not been examined yet.

In this paper, we focus on the ion-heating mechanism by standing
whistler waves.
The polarization direction of the electric field in whistler waves is
the same as the cyclotron motion of electrons. 
When the wave frequency is close to the cyclotron frequency,
$B_{\rm ext} \sim B_c$, electrons get the kinetic energy dramatically
through the resonance \cite{sano17}, and almost all the wave energy is
converted to the electrons.    
The external magnetic field considered here is larger than the
critical value, $B_{\rm ext} > B_c$, to avoid the electron cyclotron
resonance \cite{sano17}.
In the propagation of whistler waves, the stimulated Brillouin
scattering takes place, and which reduces the wave energy and drives
ion-acoustic waves \cite{nishikawa68a,forslund72,lee74}.  
However, the growth rate of the parametric decay instability is
usually much lower than the wave frequency.
In order to concentrate only on faster processes of energy conversion,
the duration of whistler waves in this analysis is limited to a few
tens of the wave periods.

We find that a substantial fraction of the wave energy is transferred
to ions as a result of the formation and immediate collapse of
standing whistler waves.
This mechanism is different from stochastic heating of underdense
plasmas by a large-amplitude standing wave \cite{hsu79,doveil81,tran82}. 
Our mechanism works only in overdense plasmas, and catalytic behavior of electron fluid is essential for the ion heating.
Hereafter, we demonstrate the ultrafast ion-heating process by
one-dimensional (1D) Particle-in-Cell (PIC) simulations. 
We also construct a theoretical model of the heating mechanism and then derive an analytical prescription of the ion temperature
achieved by the standing whistler wave heating. 
Finally, prospective applications of our heating mechanism are
discussed.  


\section{Numerical demonstration of standing whistler wave heating}

\begin{figure*}
\includegraphics[scale=0.85,clip]{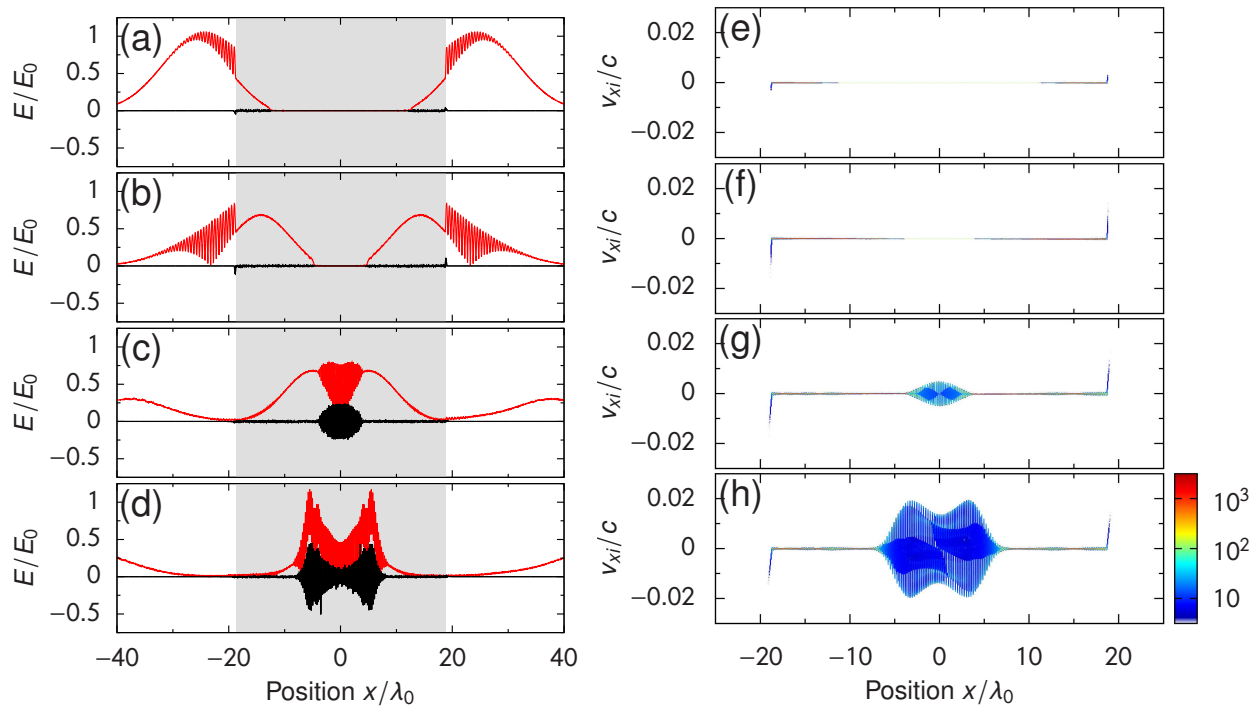}%
\caption{
Snapshots of the electric fields and ion phase diagram in the 1D PIC
simulation of the interaction between counter whistler waves and plasmas.
(a--d)
Time evolution of the electric fields during counter irradiation of
circularly polarized lights.
The longitudinal and tangential fields ($E_x$ and $|\bm{E}_{\perp}|$)
are depicted by the black and red curves, respectively.
Snapshot data are taken at (a) $\omega_0 t = -4.71$, (b) 73.8, (c)
152, and (d) 192, where the origin of time $t = 0$ is defined by the
timing when the injected lights arrive at the target surface.
The gray area stands for the inside of the target layer.
 (e--h) 
Snapshots of the position-velocity ($x$-$v_{xi}$) phase diagram for
ions taken at the same timing as in (a)-(d).
The color denotes the particle numbers.
\label{fig1}}
\end{figure*}

A simple way to form a standing electromagnetic wave is by the use of two counter beams. 
Consider a thin layer of cold hydrogen plasma in the vacuum irradiated
by CP lights with the same frequency $\omega_0$ and wavelength
$\lambda_0$ from both sides.  
In the fiducial run, the thickness of the target layer is
$\widetilde{L}_x \equiv L_x / \lambda_0 = 37.5$.
As for the initial setup, the hydrogen plasma target is located at $|x|
\le L_x / 2$ and the outside of the target is the vacuum region.
The electron density in the target is set to be overdense
$\widetilde{n}_{e0} \equiv n_{e0} / n_c = 19.3$, where $n_c =
\epsilon_0 m_e \omega_0^2 / e^2$ is the critical density, $\epsilon_0$
is the vacuum permittivity.
For simplicity, the target temperature is set to be zero initially, and we ignore the existence of the pre-plasma. 
A uniform external magnetic field is applied in the direction of the
wave propagation axis $x$ and the strength is supercritical
$\widetilde{B}_{\rm ext} \equiv B_{\rm ext} / B_c = 7.47$, which is
constant in time throughout the computation in 1D situations. 
The light traveling in the $x$ ($-x$) direction is right-hand
(left-hand) CP to the propagation direction. 
In other words, both have right-hand polarization in terms of the
magnetic field direction, and thus they enter the overdense target as
the whistler waves \cite{luan16}.
The amplitude of the incident electromagnetic wave $E_0$ is
characterized by the normalized vector potential $a_0 = e E_0 / (m_e c
\omega_0)$ where $c$ is the speed of light.
The intensity of a CP light is expressed as $I_0 = \epsilon_0 c
E_0^2$.    
A relativistic intensity with $a_0 = 2.65$ is considered and the wave
envelope shape is Gaussian with the duration of $\omega_0 \tau_0 = 
70.6$. 

The wave-plasma interaction is solved by a PIC scheme, PICLS
\cite{sentoku08}, including the Coulomb collisions.   
The escape boundary conditions for waves and particles are adopted for
both sides of the boundaries.
The CP waves are injected from both boundaries of the computational
domain, which is sufficiently broader than the target thickness.
Then the waves propagate in the vacuum for a while and then hit the target.
The transmittance and reflectivity at the target surface depends on
the refractive index of the whistler-mode $N = [1 + \widetilde{n}_{e0}
  /  (\widetilde{B}_{\rm ext} - 1) ]^{1/2}$, which is $N = 2.00$ for
the fiducial parameters.
Because the collision term is scale dependent, the physical parameters
of this run correspond to $L_x = 30$ $\mu$m, $n_{e0} = 3.37 \times
10^{22}$ cm$^{-3}$, $B_{\rm ext} = 100$ kT, $I_0 = 3 \times 10^{19}$
W/cm$^{2}$, and $\tau_0 = 30$ fs by choosing the wavelength $\lambda_0 =
0.8$ $\mu$m.
{
Here the electromagnetic wave conditions are determined based on the
typical quantities for a TW-class femtosecond laser and the target
density is equivalent to the solid hydrogen.}

The spatial and temporal resolution is $\Delta x = c \Delta t =
\lambda_0 / 10^3$ and the particle number is 200 per each grid cell at the
beginning.
In the strongly magnetized plasmas, the time resolution $\Delta t$
should be shorter than the electron gyration time as well as the
plasma oscillation time.
Otherwise, the unphysical numerical heating breaks the energy
conservation.
In order to capture the propagation of the whistler waves and the
evolution of the standing waves correctly, it would be better for the
whistler wavelength to be resolved by a few hundreds of grid cells. 
These conditions are satisfied in all simulations shown in this paper.
We have confirmed by the convergence check that the conclusions
discussed in our analysis are unaffected by the numerical resolution.

\begin{figure*}
\includegraphics[scale=0.85,clip]{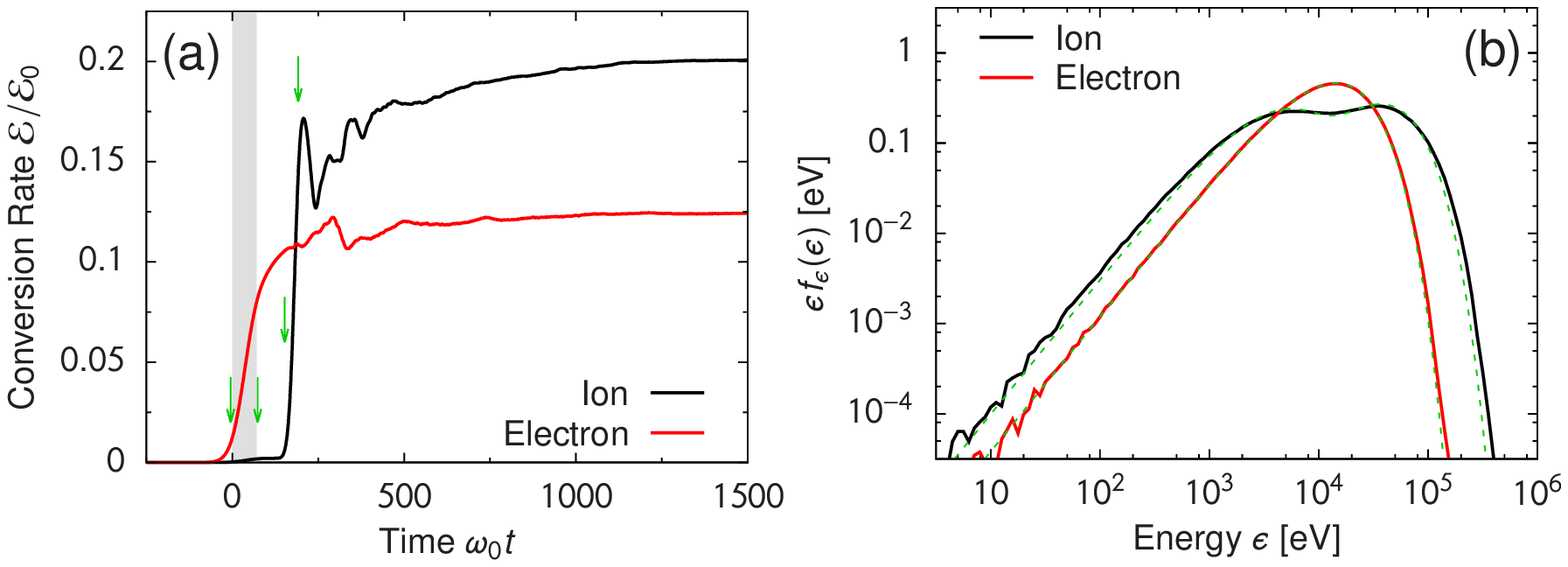}%
\caption{
Energy conversion rate and energy spectra for ions and electrons in the
fiducial run.
(a)
Time histories of the ion and electron energies normalized by the
injected energy of the electromagnetic wave ${\cal{E}}_0$.
The gray color in the figure stands for the pulse duration of the
target irradiation, $0 \le \omega_0 t \le 70.6$. 
The arrows indicate the snapshot timing shown in Fig.~\ref{fig1}.
(b)
Energy spectra for the ions and electrons at the end of the
calculation.
The vertical axis is $\epsilon f_{\epsilon} (\epsilon)$ where
$f_{\epsilon} (\epsilon)$ is the probability density function for the
energy $\epsilon$, so that the peak value is related to the
temperature as $\epsilon_{\rm peak} = (3/2) k_{B} T$ in the
Maxwell-Boltzmann distribution.
The ion spectrum is fitted well by two temperature model with $T_i = 24$ and $2.9$ keV, and the electron spectrum is almost identical to the
thermal distribution of $T_e = 9.6$ keV.
The dotted curves show the fitted functions.
\label{fig2}}
\end{figure*}

The PIC simulation clearly shows that ions in the overdense target are heated efficiently by the counter irradiation of CP lights.
Because of the no-cutoff feature, both of the counter beams propagate
inside of the target as whistler waves. 
When the two whistler waves pass each other, a standing whistler wave
is formed in the middle of the target (Fig.~\ref{fig1}).
Right after the appearance of the standing wave, the longitudinal
electric field $E_x$ is generated to the similar order of the
transverse wave field $E_0$. 
At the same time, ions start to be accelerated by the electric field
$E_x$.
The ion velocity increases quickly up to 2\% of the light speed within
several wave periods.  
The rapid increase of ion energy can be seen in Fig.~\ref{fig2}(a)
where temporal histories of the total amount of energies for ions and electrons are depicted.
When the injected wave fields step into the target layer, only
electrons start to move due to the quiver motions of the whistler
waves, whereas ion motion exhibits little change.  
However, the ion energy is jumped up at the timing when two whistler
waves are overlapped at the center of the target, and ultimately the
ion energy exceeds the electron energy.
One-third of the wave energy is absorbed by plasmas through this
interaction, where the ions gain more than 60\% of the absorbed
energy, i.e., the conversion efficiency from the waves to ions is
$\sim$ 20\%.
The acquired ion energy is drastically enhanced by an order of
magnitude compared with the case without the external magnetic field,
or no whistler-mode case. 
 
The energy spectrum of ions is nearly thermalized at the later stage
far beyond the pulse duration, $\omega_0 t_{\rm end} = 2.78 \times
10^3 \gg \omega_0 \tau_0$ [Fig.~\ref{fig2}(b)]. 
The peak energy of the ion spectrum corresponds to about $T_i \sim 24$
keV, which is higher than the electron temperature of $T_e \sim 9.6$ keV.
{
The energy density of the external magnetic field is still larger than
that of the thermalized plasma.
The plasma beta is about 0.02 at the end of calculation for this case.}

This series of events is the ion-heating scenario by standing whistler
waves in the numerical simulation.
Surprisingly, thermal ion plasma over tens of keV in solid density is
produced by Joule-class lasers in a rather simple geometry only if a
sufficiently strong magnetic field is available.

\section{Theoretical modeling}

Next, we will give a theoretical model of the ion-heating mechanism
based on fundamental equations.  
The relativistic effects are neglected in the
following analytical discussion. 
Eventually, it turns out that the ion temperature heated by counter
whistler waves is described by a simple formula of the initial
wave-plasma conditions.  

The eigen functions of the tangential electric field $\bm{E}_w$, magnetic
field $\bm{B}_w$ and electron velocity $\bm{v}_w$ for counter whistler
waves traveling in the $\pm x$ direction with the wavenumber $k_w
\equiv 2 \pi / \lambda_w$ are given by  
\begin{eqnarray}
\bm{E}_w^{\pm} &=& E_w^{\pm} \exp
\left[ i \left( \pm k_w x - \omega_0 t \right) \right] 
(\widehat{y} + i \widehat{z} ) \;, \label{eq:ew} \\
\bm{B}_w^{\pm} &=& \mp B_w^{\pm} \exp [ i ( \pm k_w x - \omega_0 t) ] 
(i \widehat{y} - \widehat{z} )  \;, \label{eq:bw} \\
\bm{v}_w^{\pm} &=& v_w^{\pm} \exp [ i ( \pm k_w x - \omega_0 t) ] 
(i \widehat{y} - \widehat{z} )  \;, \label{eq:vw}
\end{eqnarray}
where $B_w^{\pm} = ({k_w}/{\omega_0}) E_w^{\pm}$ and
\begin{equation}
v_w^{\pm} = \frac{1}{\widetilde{B}_{\rm ext} -1}\frac{e}{m_e \omega_0} E_w^{\pm}
\;.
\label{eq:vw2}
\end{equation}
Suppose both of the injected CP lights have the same wavenumber $k_0$
and amplitude $a_0$ in the vacuum, the transmitted whistler
waves will have $k_w = N k_0$ and $a_w = 2 a_0 / (N+1)$ \cite{luan16}.
Since $N$ is larger than unity when $B_{\rm ext}$ is supercritical,
the wavelength and amplitude of the electromagnetic waves become
shorter and smaller in the target. 
The nonrelativistic condition is then given by $a_w <
\widetilde{B}_{\rm ext} - 1$ from Eq.~(\ref{eq:vw2}).

Let us consider the force balance for the electron fluid in the
longitudinal direction.   
The electromotive force, $- e (\bm{v}_e \bm{\times} \bm{B})_x$,
applied to the electrons is always zero, that is, $(\bm{v}_w^{\pm}
\bm{\times} \bm{B}_w^{\pm} )_x = 0$ for the case of a single whistler
wave.
However, this term becomes finite in the standing whistler wave,
$[ (\bm{v}_w^{+} + \bm{v}_w^{-}) \bm{\times} (\bm{B}_w^{+} +
  \bm{B}_w^{-}) ]_x \ne 0$, 
which brings curious consequence in the evolution of plasmas located
at the standing wave.  

\begin{figure*}
\includegraphics[scale=0.85,clip]{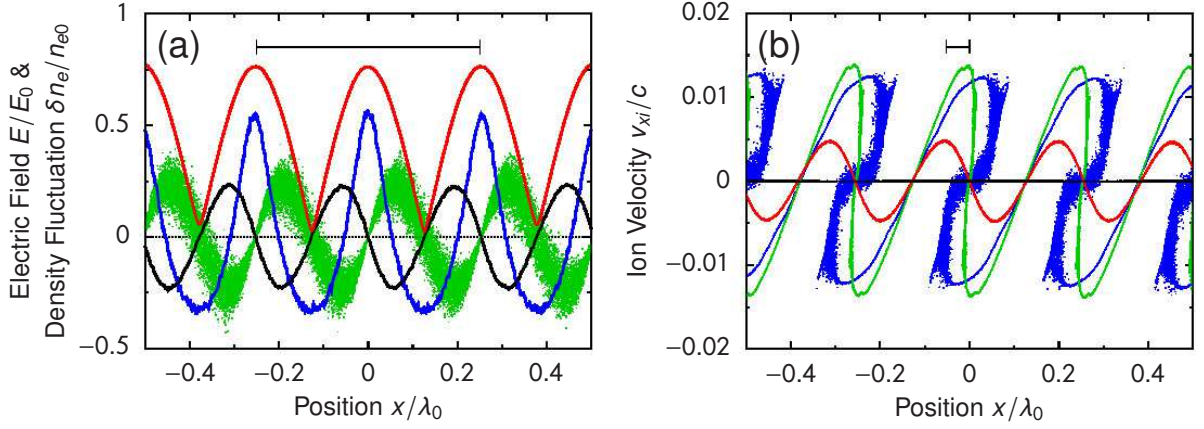}%
\caption{
Detailed structures of a standing whistler wave and linearly
growing velocity fluctuations of ions in the fiducial run.
(a)
Magnified view of the electric fields at the center of the target,
$E_x$ (black) and $|\bm{E}_{\perp}|$ (red), taken at $\omega_0 t = 152$,
which is right after the formation of the standing whistler wave.
The electromotive field $(\bm{v}_e \bm{\times} \bm{B})_x$ worked on the
electrons is also plotted by green dots.
The blue curve denotes the electron density fluctuation $\delta n_e
\equiv n_e - n_{e0}$.
The length indicated in the figure corresponds to the wavelength
of the whistler wave $\lambda_w = \lambda_0 / N$.
(b)
Amplitude growth of the ion velocity in a standing wave identified
from four successive snapshots at $\omega_0 t = 113$ (black), 152
(red), 192 (green), and 231 (blue).
The indicated length scale is $\lambda_w/8$.
\label{fig3}}
\end{figure*}

\begin{figure*}
\includegraphics[scale=0.85,clip]{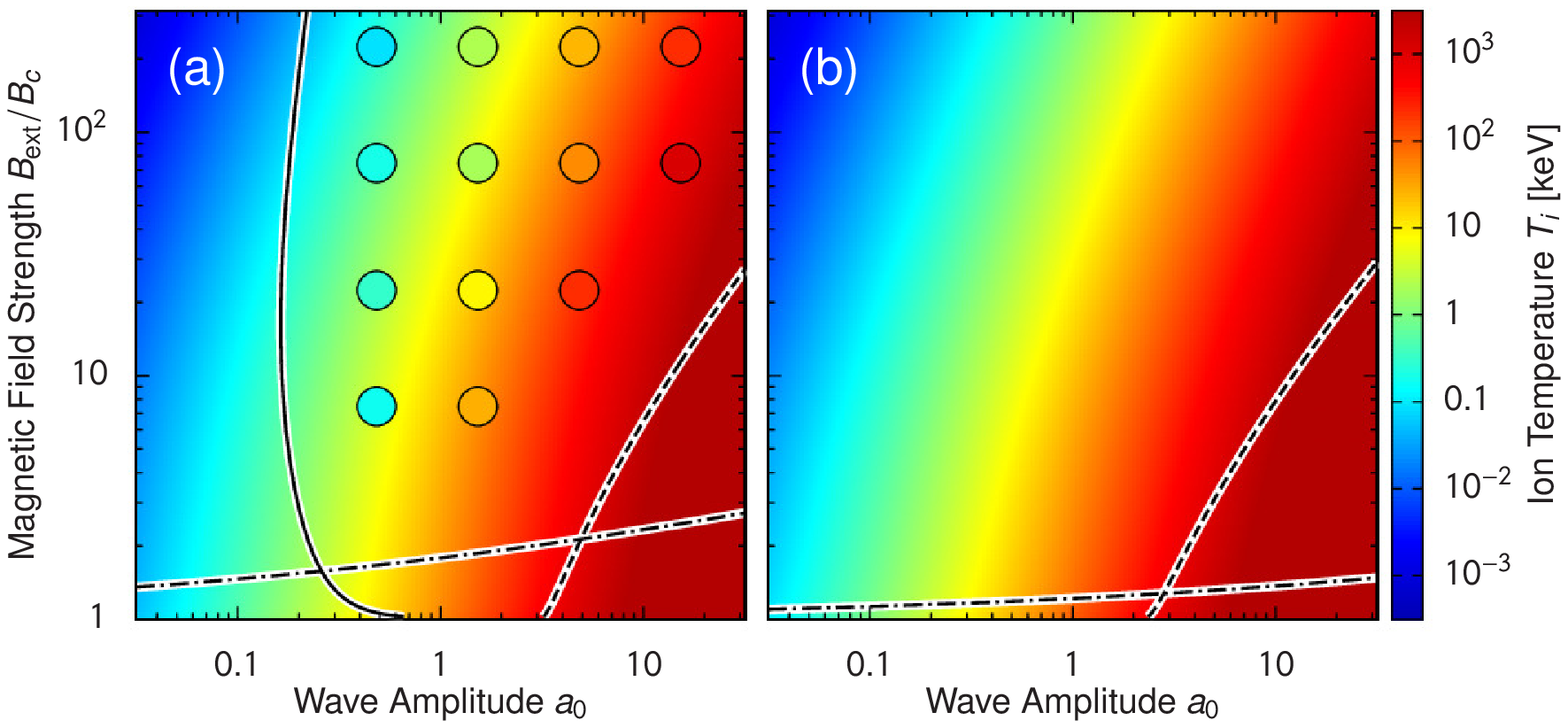}%
\caption{
Dependence of the model prediction for the ion temperature on the initial
parameters $a_0$, $B_{\rm ext} / B_c$, and $n_e / n_c$.
(a)
Predicted ion temperature given by Eq.~(\ref{eq:ti}) for the cases
of $n_{e0} / n_c = 19.3$.
The wavelength is assumed to be $\lambda_0 = 0.8$ $\mu$m.
Overplotted colored circles indicate $T_i$ obtained by the 1D PIC
simulations with given parameters $a_0$ and $B_{\rm ext} / B_c$.
In these runs, a flat-top pulse shape is used with a sufficiently long
duration $\tau_0 \gg \tau_{\rm sat}$.
The valid range of the temperature prediction is within the area
surrounded by three critical curves, which are obtained from the
pressure-gradient condition [Eq.~(\ref{eq:dp}); solid curve] and the cyclotron
resonance conditions [Eqs.~(\ref{eq:cr1}) and (\ref{eq:cr2}); dashed and
dot-dashed curves].
(b)
Same diagram for the cases of $n_{e0} / n_c = 10$ and $\lambda_0 = 1$
cm.
\label{fig4}}
\end{figure*}

The force free condition, $E_x + (\bm{v}_e \bm{\times} \bm{B})_x
\approx 0$, should be satisfied for the electron fluid, because the
inertial and pressure-gradient terms in the electron equation of 
motion are negligible in the current situation ($\widetilde{n}_{e0}
\gg 1$ and $T_e \sim 0$). 
The force balance for electrons is established distinctly in
the PIC simulation [Fig.~\ref{fig3}(a)].
The electromotive force acts as the negative ponderomotive force in
this circumstance.
The electrons gather periodically at the antinodes of the standing
whistler waves to generate the longitudinal electric field $E_x$.
The amplitude of $E_x$ is then evaluated as
\begin{equation}
E_x \approx - \left[ \frac{8 N a_0^2}{(N + 1)^2
(\widetilde{B}_{\rm ext} - 1)} \frac{m_e c \omega_0}{e}
\right] \sin (2 N k_0 x) \;,
\label{eq:ex}
\end{equation}
by using Eqs. (\ref{eq:bw}) and (\ref{eq:vw}), which has a sinusoidal
distribution with the wavelength of $\lambda_w/2$. 
Interestingly, the electric field $E_x$ is constant in time, so that
the ions are accelerated effectively by this tiny-scale steady force.  

\subsection{Ion temperature}

The equation of motion for ions is approximately written as
$\partial v_{xi} / \partial t \approx Z e E_x / m_i$,
because of the slow velocity and low temperature initially ($v_{xi}
\ll c$ and $T_i \sim 0$). 
Here $m_i$ and $Z$ are the ion mass and charge number, respectively.
Since the electric force is independent of time, the amplitude of
ion velocity increases linearly with time,
\begin{equation}
v_{xi} \approx \frac{Z e E_x}{m_i} (t - t_s) \;,
\end{equation}
where the displacement of the ion position is ignored, and $t_s$ is
the time when the standing whistler wave appears in the target. 
As seen from Fig.~\ref{fig3}(b), the constant acceleration of ion
velocity is consistent with the PIC results from $\omega_0 t \sim 120$ to 190. 
The ion density increases at the antinodes of the standing wave in the
same manner as the electrons. 
Then the amplitude of the periodic density fluctuation of the
electrons increases even further instantaneously to sustain the
constant electric field $E_x$ as long as the standing waves survive.  
This positive feedback cycle continues to accelerate ions.

The ion acceleration will be terminated by the steepening of the
waveform in the position-velocity phase diagram.
Due to the huge density fluctuation caused by the localization of
electrons and ions, the standing wave is no longer sustained and
breaks down at the saturation time $\tau_{\rm sat}$ that is approximately
estimated by
\begin{equation}
\int^{t_s + \tau_{\rm sat}}_{t_s} | v_{xi} | dt \sim \frac{\lambda_w}8 \;,
\end{equation}
{
where $\lambda_w / 8$ corresponds to the acceleration length for the
fastest ions, that is, a quarter of the ion wavelength [see
  Fig.~\ref{fig3}(b)].} 
Solving this relation, the maximum amplitude is obtained as
\begin{equation}
\frac{v_{xi,\max}}{c} \sim
\left[ \frac{4 \pi a_0^2}{(N + 1)^2 (\widetilde{B}_{\rm ext} - 1)}
\frac{Z m_e}{m_i} \right]^{1/2} \;,
\end{equation}
at the time 
\begin{equation}
\omega_0 \tau_{\rm sat} \sim
\left[
  \frac{\pi}{16}
  \frac{(N + 1)^2 (\widetilde{B}_{\rm ext} - 1)}
{N^2 a_0^2} \frac{m_i}{Z m_e} \right]^{1/2} \;.
\end{equation}
The solution suggests that the wave duration $\tau_0$ must be longer 
than $\tau_{\rm sat}$ in order for ions to gain the maximum energy from the
standing whistler wave.
The saturation time could be a few tens of the wave period or even
shorter for the relativistic intensity cases $a_0 \gtrsim 1$.

After the steepening, counter ion flows coexist at the same location, and the ions begin to thermalize through wave breaking and
kinetic instabilities like ion two-stream instability \cite{ross13}.
If the accelerated ions are totally thermalized, it will give a
reasonable evaluation of the maximum ion temperature, i.e., 
\begin{equation}
\frac{k_B T_i}{m_e c^2} \sim 
\frac{2 \pi}3 \frac{a_0^2 Z}
{(N + 1)^2 (\widetilde{B}_{\rm ext} - 1)} \;,
\label{eq:ti}
\end{equation}  
where the relation
$\langle v_i^2 \rangle = {3 k_B T_i}/{m_i} \sim v_{xi,\max}^2/2$
is adopted. 
The PIC simulations confirm that the modeled temperature calculated
from Eq.~(\ref{eq:ti}) is genuinely reliable to interpret the outcome of
the counter CP light irradiation [see Fig.~\ref{fig4}(a)]. 
According to the theoretical model, the final ion temperature is
independent of the ion mass, but proportional to the charge $Z$.
In the overdense limit, $\widetilde{n}_{e0} \gg \widetilde{B}_{\rm ext}
> 1$, the dependence of $T_i$ is proportional to $a_0^2 Z /
\widetilde{n}_{e0}$.
During the collapsing regime, only ions are accelerated and heated
selectively. 
That is why it can be regarded as a mechanism of direct ion-heating by
electromagnetic waves. 
Note that the same phenomenon occurs by the left-hand CP lights if the
plasma density is overdense and less than the L-cutoff, $1
\lesssim \widetilde{n}_{e0} \lesssim \widetilde{n}_L \equiv
\widetilde{B}_{\rm ext} + 1$.
The standing waves could be generated even with a single whistler wave
by the reflection at the rear edge of a thin target.

\subsection{Electron heating}

The electron heating, on the other hand, would be dominated by the
resistive heating at least in the 1D situation.
The energy equation is given by
\begin{equation}
\frac32 \frac{\partial}{\partial t} (k_B T_e) \approx 
m_e \nu_{ei} | \bm{v}_e - \bm{v}_i |^2 \;,
\end{equation}  
where the relative velocity between electrons and ions is mainly
caused by the quiver motion of whistler waves [Eq. (\ref{eq:vw2})]. 
Assuming the Maxwellian-averaged collision frequency \cite{kruer03,chen84},
\begin{equation}
  \nu_{ei} =
  \frac{\ln \Lambda}{3 (2 \pi)^{3/2}}
  \frac{Z e^4}{\varepsilon_0^2 m_e^{1/2}}
  \frac{n_{e}}{(k_B T_e )^{3/2}} \;,
\end{equation}  
the electron temperature is derived as
\begin{equation}
\frac{k_B T_e}{m_e c^2} \sim
\left[
\frac{40 \sqrt{2 \pi} \ln \Lambda}{9} \frac{a_0^2 \widetilde{n}_{e0}}
{(N+1)^2 (\widetilde{B}_{\rm ext} -1)^2}
\frac{Z r_e}{\lambda_0} \omega_0 t
\right]^{2/5} \;,
\label{eq:te}
\end{equation}  
where
$r_e = {e^2}/({4 \pi \varepsilon_0 m_e c^2})$ is the electron
classical radius. 
There is a wide parameter range where the ion temperature $T_i$
becomes higher than $T_e$.

{
In the PIC simulations, we neglect the initial temperature of the
target. 
Even when the finite temperature is considered initially, the
ion-heating process is found to be unchanged if the initial electron
temperature is lower than the temperature given by Eq.~(\ref{eq:te}).} 

\subsection{Valid range of the model prediction}

The ion temperature is now easily estimated from three initial
parameters ($a_0$, $\widetilde{B}_{\rm ext}$, and $\widetilde{n}_{e0}$)
with the help of Eq.~(\ref{eq:ti}).
Figure~\ref{fig4}(a) shows the predicted ion temperature for the cases
of $\widetilde{n}_{e0} = 19.3$ assuming a typical wavelength of
high-intensity lasers $\lambda_0 = 0.8$ $\mu$m.
Numerically obtained ion temperatures in the PIC simulations are
overplotted by the colored circles, which show good agreement with the
model prediction in a wide range from $T_i \sim 100$ eV to 1 MeV.
The deviation from the model prediction is within a factor of 2 for
the cases of $a_0 \gtrsim 1$.

It should be noticed that there is a valid range of the theoretical
model. 
One of the essential quantities of this heating mechanism is the
longitudinal field $E_x$ given by Eq.~(\ref{eq:ex}).   
If the pressure-gradient term in the electron equation of motion is not negligible, the electromotive force could balance with $\nabla
P_e$, and then the static electric field would not appear.
Therefore, our model requires 
$\nabla P_e \sim 2 k_w P_e \lesssim e n_e (\bm{v}_e \bm{\times}
\bm{B})_x$, which is rewritten by the initial parameters as 
\begin{equation}
a_0 \gtrsim \left[ \frac{5 \sqrt{2 \pi} \ln \Lambda}{36} 
(N + 1)^3 (\widetilde{B}_{\rm ext} - 1)^{1/2}
\widetilde{n}_{e0} \frac{Z r_e}{\lambda_0} \omega_0 t
\right]^{1/3} \;,
\label{eq:dp}
\end{equation}
where $\ln \Lambda$ is the Coulomb logarithm.
This validity condition is shown by the solid curve in Fig.~\ref{fig4}
assuming $\ln \Lambda = 10$, $Z = 1$, and $\omega_0 t = 300$.
The pressure-gradient term is dominant for the lower intensity cases,
because of the weaker dependence $\nabla P_e \propto T_e \propto a_0^{4/5}$
than $(\bm{v}_e \bm{\times} \bm{B})_x \propto a_0^2$.

Another requirement is to avoid the electron cyclotron resonance,
which prevents whistler wave propagation by disturbing the electron
quiver motion.
{
The relativistic and Doppler effects must be considered to derive the
resonance condition, $\omega_0 - k_w v_{\parallel} = \omega_{ce}/\gamma$
\cite{sano17}.
Assuming $v_{\parallel}$ is of the order of the thermal velocity
$v_{\rm th}$, the resonance condition is summarized as
$\widetilde{B}_{\rm ext} \gtrsim \gamma$ and $\widetilde{B}_{\rm ext}
\gtrsim 1 + \widetilde{n}_{e0}^{1/3} (v_{\rm th} / c)^{2/3}$
in the relativistic and nonrelativistic limit, respectively. 
Here $\gamma \sim (1 + a_w^2)^{1/2}$ and $v_{\rm th} = (k_B T_e /
m_e)^{1/2}$ are the Lorentz factor and thermal velocity of electrons.}
By using the initial parameters, these conditions are settled in
\begin{equation}
a_0 \lesssim \frac{(N + 1)(\widetilde{B}_{\rm ext}^2 -1 )^{1/2}}2 \;,
\label{eq:cr1}
\end{equation}  
and
\begin{equation}
\widetilde{B}_{\rm ext} \gtrsim 1 + 
\left[ \frac{40 \sqrt{2 \pi} \ln \Lambda}9
\frac{a_0^2 \widetilde{n}_{e0}^{7/2}}{(N + 1)^2} 
\frac{Z r_e}{\lambda_0} \omega_0 t
\right]^{2/19} \;,
\label{eq:cr2}
\end{equation}
which are also plotted by the dashed and dot-dashed curves in
Fig.~\ref{fig4}.
In the end, the ion temperature given by Eq.~(\ref{eq:ti}) is
applicable only within the area surrounded by these three curves.  

\section{Parameter dependence of ion energy increase}

\begin{figure}
\includegraphics[scale=0.85,clip]{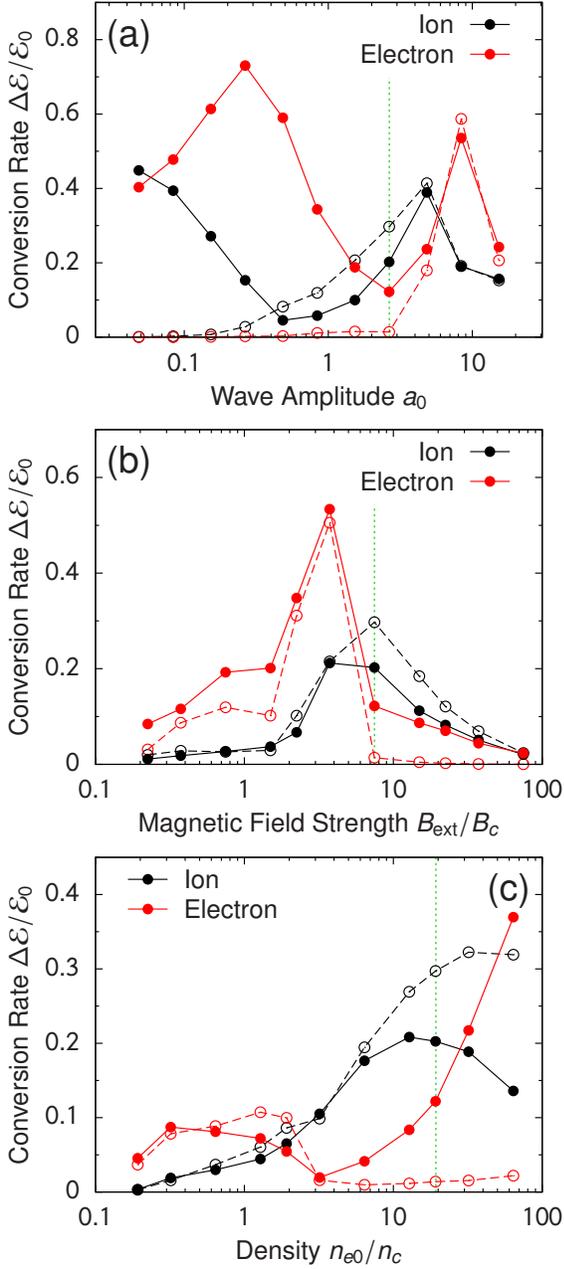}%
\caption{
Parameter dependence of the energy conversion rate for ions and
electrons obtained by 1D PIC simulations.
(a) The conversion rate for ions (black) and electrons (red) as a
function of the wave amplitude $a_0$.
The conversion rate is evaluated at $\omega_0 t_{\rm end} = 2.78
\times 10^3$.
The filled and open circles denote the results with and without
the Coulomb collision effects, respectively.
The vertical dotted line indicates the fiducial parameter $a_0 =
2.65$.
(b and c) Dependence on (b) the external field strength $B_{\rm ext} /
B_c$ and (c) the plasma density of the target $n_{e0} / n_c$.
The meanings of the marks are the same as in (a).
The initial parameters are the same as in the fiducial run except
for (b) $B_{\rm ext}$ and (c) $n_{e0}$.
The fiducial parameters are $B_{\rm ext} / B_c = 7.47$ and $n_{e0} /
n_c = 19.3$.
\label{fig5}}
\end{figure}

When the standing whistler wave heating sets in, the total ion energy
is usually higher than the electron energy.
Then the energy conversion rate will be a good indicator of the ion
heating by standing whistler waves.
Figure~\ref{fig5} shows the parameter dependence of the energy conversion rate for ions and electrons obtained from 1D PIC simulations
similar to the fiducial run.

First, we investigate the effects of the incident wave amplitude.
As seen from Fig.~\ref{fig5}(a), where the initial parameters of these
runs are identical to those in the fiducial run except for $a_0$, the
ion energy is dominant when the wave amplitude is about $a_0 \sim
1$--$5$. 
In other cases, the formation of standing waves is inhibited by
higher electron temperature by the resistive heating or the cyclotron
resonance.
This is actually consistent with the validity conditions given by
Eqs.~(\ref{eq:dp}) and (\ref{eq:cr1}).

The next parameter is the strength of the external magnetic field.
As expected, the ion energy is dominant only when the magnetic
field $B_{\rm ext}$ is sufficiently larger than the critical value
$B_c$ [Fig.~\ref{fig5}(b)].
The conversion efficiency decreases as the field strength increases 
so that the best condition for the ion heating is around
$\widetilde{B}_{\rm ext} \sim$ 5--10. 

As for the target density, obviously it must be overdense
$\widetilde{n}_{e0} \gtrsim 2$ for efficient ion heating
[Fig.~\ref{fig5}(c)]. 
In the overdense limit,
the electron temperature given by Eq.~(\ref{eq:te}) is
independent of the initial density $n_{e0}$, while the ion temperature
has a dependence $T_i \propto n_{e0}^{-1}$.
Then the energy fraction of electrons becomes predominant in this limit.

The wavelength of the incident CP lights is assumed to be $\lambda_0 =
0.8$ $\mu$m in our PIC simulations.
The Coulomb collision term in the equation of motion for the charged particles has a dependence on the critical density.
If the longer wavelength is selected, the relative importance of the collision effects becomes weaker.
In Fig.~\ref{fig5}, the results of collisionless simulations are also
shown as a reference.
When the collision effect is negligible, or the wave frequency is
sufficiently high, the energy conversion to electrons is reduced
significantly.
Figure~\ref{fig4}(b) indicates the ion temperature heated with the
whistler wavelength of $\lambda_0 = 1$ cm assuming the target density
$\widetilde{n}_{e0} = 10$. 
The validity curves are largely different from those in the $\lambda_0
= 0.8$ $\mu$m cases.
The pressure-gradient condition given by Eq.~(\ref{eq:dp}) is out of
range in this figure ($a_0 < 0.01$).
Therefore the standing whistler wave heating is realized in a broader range of the plasma parameters.  

\section{Discussion and Conclusions}

\begin{table*}
  \begin{tabular}{lccccccc}
    \hline \hline
& \multicolumn{4}{c}{Frequency (Wavelength)} & & & \\ \cline{2-5}    
& \begin{tabular}{c} 300 THz \\ (1 $\mu$m) \end{tabular} &
\begin{tabular}{c} 30 THz \\ (10 $\mu$m) \end{tabular} &
\begin{tabular}{c} 30 GHz \\ (1 cm) \end{tabular} &
\begin{tabular}{c} 3 kHz \\ (100 km) \end{tabular} & &
Parameter & Range \\
\cline{1-5}\cline{7-8}
\begin{tabular}{l} Wave Amplitude \\ (W/cm$^2$) \end{tabular} & 
$3 \times 10^{18}$--$10^{20}$ & $3 \times 10^{16}$--$10^{18}$ &
$3 \times 10^{10}$--$10^{12}$ & $3 \times 10^{-4}$--$10^{-2}$ & &
$a_0$ & $1$--$5$ \\
\begin{tabular}{l} Magnetic Field \\ Strength (T) \end{tabular} & 
$5 \times 10^4$--$10^5$ &    $5 \times 10^3$--$10^4$ &
$5$--$10$ & $5 \times 10^{-7}$--$10^{-6}$ & &
$B_{\rm ext}/B_c$ & $5$--$10$ \\
\begin{tabular}{l} Density \\ (cm$^{-3}$) \end{tabular} & 
$2 \times 10^{21}$--$10^{23}$ & $2 \times 10^{19}$--$10^{21}$ &
$2 \times 10^{13}$--$10^{15}$ & $0.2$--$10$ & &
$n_{e0}/n_c$ & $2$--$100$ \\
Application & glass \& TiSap laser & CO$_2$ laser & tokamak &
\begin{tabular}{c} planetary \\ magnetosphere \end{tabular} 
\\    \hline \hline
  \end{tabular}
\caption{
Characteristic physical quantities for the thermal ion-plasma
generation over 10 keV.
The appropriate values of the wave amplitude, external magnetic field
strength, and plasma density are listed for various cases of the
frequency (wavelength) of electromagnetic waves. 
The range of each quantities in the non-dimensional parameter are also
listed in the table.
\label{tab1}}
\end{table*}

The standing whistler wave heating is expected to develop various applications.
For ICF, ions should be heated up to the higher temperature exceeding keV in imploded dense plasmas. 
Our method might give an advanced technique for an alternative
ignition scheme of ICF by a completely different use of magnetic
fields from the previous ideas
\cite{hohenberger12,perkins13,wang15,fujioka16,sakata18}.  
The keV ion plasma generated by this method could be an efficient
thermal neutron source \cite{baker18}.
Since it requires the existence of a strong magnetic field larger than
$B_c \approx 10$ kT for $\lambda_0 = 1$ $\mu$m, practically the
generation of such an extreme magnetic field would be the first serious barrier to be resolved.    
Recently, the achievement of strong magnetic fields of
kilo-Tesla order in laser experiments has been reported by several
groups \cite{yoneda12,fujioka13,goyon17,santos18}. 
Then it would be plausible in the near future to excite
relativistic whistler waves from high-intensity lasers under a
supercritical field condition $B_{\rm ext} > B_c$ \cite{korneev15}. 

The critical value $B_c$ can be reduced significantly by a choice of
the longer wavelength.
The typical quantities suitable for the standing whistler wave heating
are summarized in Table 1. 
The carbon dioxide laser of the wavelength $\lambda_0 = 10$ $\mu$m
might be a better choice for the proof-of-principle experiment of this mechanism, because the critical field strength decreases by an order
of magnitude.
If the wavelength is of the order of centimeter, the critical field
strength goes down to $B_c \sim 1$ T.
The situation shown in Figure~\ref{fig4}(b) corresponds to a tokamak
plasma of the density $n_{e0} = 1.11 \times 10^{14}$ cm$^{-3}$
($\widetilde{n}_{e0} = 10$) when the wavelength $\lambda_0 = 1$ cm is
used. 
Based on the model prediction, the intensity $a_0 \sim 0.5$ is
needed to produce 10 keV ion plasma under a ITER-relevant magnetic
field ($B_{\rm ext} \sim 5$ T) \cite{stacey05}. 
The other extreme case is $\lambda_0 \sim 100$ km, or $\omega_0 / (2
\pi) \sim 3$ kHz, which gives $B_c \sim 100$ nT.
These quantities are appropriate to the ion acceleration in
planetary magnetospheres \cite{bagenal92}.
It must be meaningful to pursue various applicability of this
mechanism by a series of PIC simulations. 

{
Ions and electrons are heated by the decay of whistler turbulence
observed in the solar wind \cite{alexandrova13,hughes14,gary16}.
Interactions of counter-propagating waves should frequently occur in
the turbulence so that it is interesting to
examine the collisions of two waves with different frequencies.
In this study, after the collapse of the standing whistler wave,
the turbulent state of low-beta plasma is excited by ion kinetic
instabilities and residual whistler waves. 
The gradual increase of the ion energy after the wave breaking [see
Fig.~\ref{fig2}(a)] might be caused by decaying whistler turbulence.
Thus the multi-dimensional study of the turbulent stage would be
applicable to the solar wind problem.}


In summary, an ion-heating mechanism by the counter configuration of
whistler waves has been investigated numerically and theoretically.   
The critical process is the collapse of standing whistler waves, which
enables direct energy transfer from the electromagnetic waves to
ions. 
The ion temperature is found to be estimated very accurately from
three initial parameters that are the wave amplitude $a_0$, magnetic
field strength $\widetilde{B}_{\rm ext}$, and plasma density
$\widetilde{n}_{e0}$. 
Typical parameter ranges for thermal plasma generation over 10 keV are
$a_0 \sim$ 1--5, $\widetilde{B}_{\rm ext} \sim$ 5--10, and
$\widetilde{n}_{e0} \sim$ 2--100.  
If a pair of linearly polarized lights are used instead of the counter CP lights, a part of the incident lights is converted to the whistler-mode and enters the overdense target \cite{sano17}.
Thus the same mechanism of ion heating takes place by the transmitted
whistler waves, but the energy conversion efficiency is much lower
than the CP cases. 

Although we focus on 1D results in this paper, 2D PIC simulations,
where the periodic boundary condition is imposed in another spatial
direction $y$, reveals successfully that the same ion heating occurs
in 2D as well (see Appendix).  
The only difference is observed in the higher electron temperature than the 1D counterpart when the width of the computational domain in
the $y$ direction becomes comparable to the whistler wavelength.
The detailed analysis in multi-dimensional cases will be an essential subject for our future work.

\begin{acknowledgments}
We thank Shinsuke Fujioka, Masahiro Hoshino, Natsumi Iwata, Yasuaki
Kishimoto, and Youichi Sakawa for fruitful discussions.
This work was supported by JSPS KAKENHI Grant Numbers JP15K17798,
JP15K21767, JP16H02245, and JP17J02020 and by JSPS Core-to-Core
Program Asia-Africa Science Platform ``Research and Education Center
for Laser Astrophysics in Asia''.
\end{acknowledgments}


\appendix*

\section{Two-dimensional effects}

\begin{figure*}
\includegraphics[scale=0.85,clip]{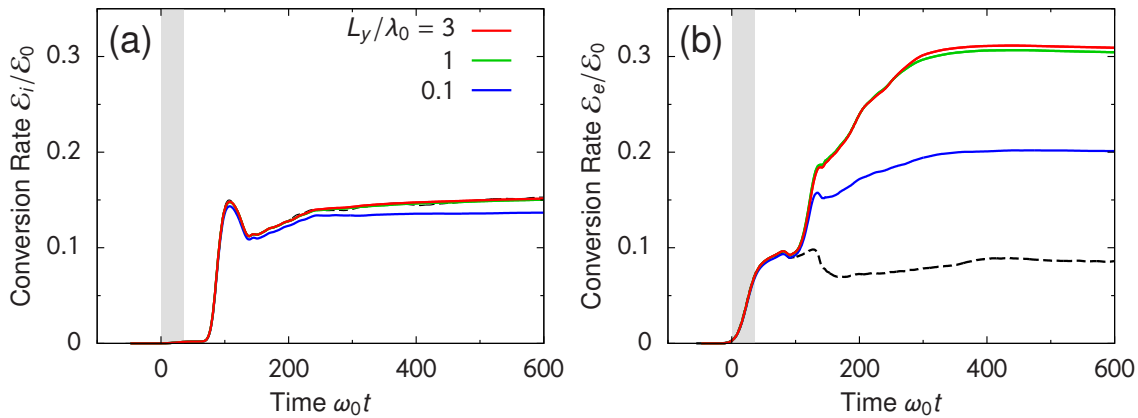}%
\caption{
Time evolution of energy conversion rate for (a) ions
and (b) electrons is depicted for various cases of $L_y / \lambda_0 = 3$
(red), 1 (green), and 0.1 (blue) of the 2D runs.
The initial parameters are the same as in the 1D fiducial run except
for the target thickness $L_x / \lambda_0 = 18.75$ and the pulse
duration $\omega_0 \tau_0 = 35.3$.
The dashed curve is the result of the 1D run, that is, $L_y = 0$.
\label{fig6}}
\end{figure*}

\begin{figure}
\includegraphics[scale=0.85,clip]{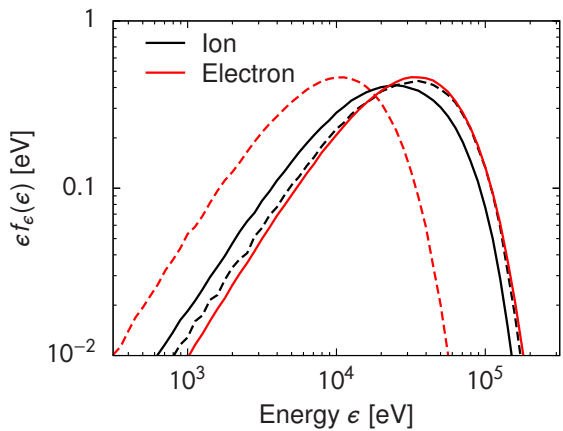}%
\caption{
Energy spectrum $\epsilon f_{\epsilon} (\epsilon)$ of ions (black)
and electrons (red) at the end of the calculation $\omega_0 t_{\rm
  end} = 1.84 \times 10^{3}$ are plotted for the cases of $L_y /
\lambda_0 = 3$ (solid) and 0 (bashed).
These spectra are calculated from the particles located at the
standing wave region $|x/\lambda_0| \le 5$.
\label{fig7}}
\end{figure}

\begin{figure*}
\includegraphics[scale=0.85,clip]{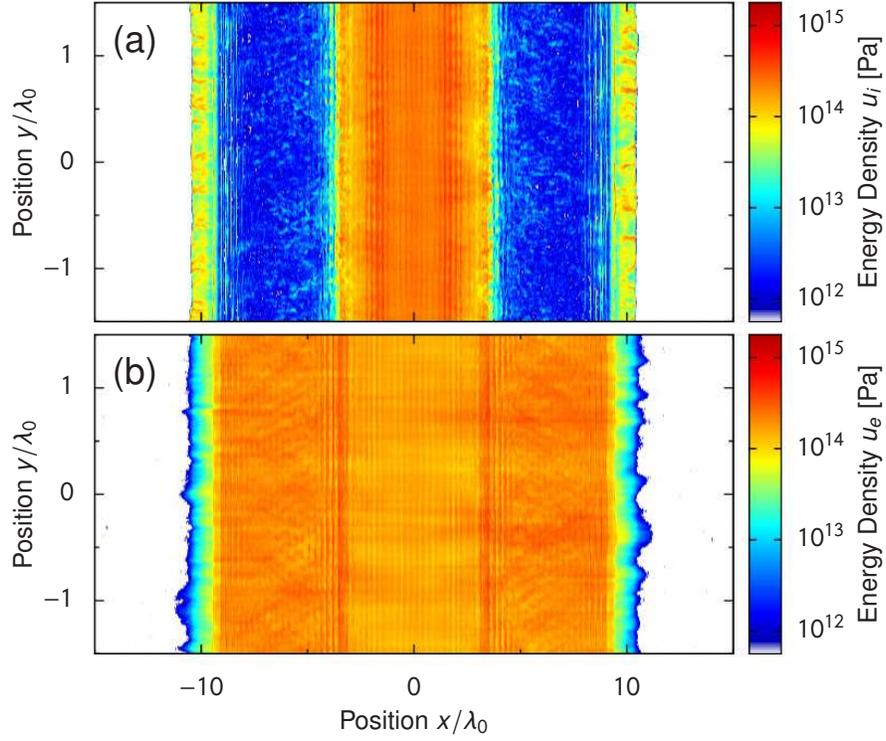}%
\caption{
Spatial distributions of the energy density $u$ for (a) ions and
(b) electrons in a 2D simulation.
The energy density is measured in the unit of Pascal.
The initial parameters are the same as in the 1D fiducial run except
for the target thickness $L_x / \lambda_0 = 18.75$ and the pulse
duration $\omega_0 \tau_0 = 35.3$.
The box size in the $y$ direction is $L_y / \lambda_0 = 3$.
These images are taken at $\omega_0 t = 329$.
\label{fig8}}
\end{figure*}

\begin{figure*}
\includegraphics[scale=0.85,clip]{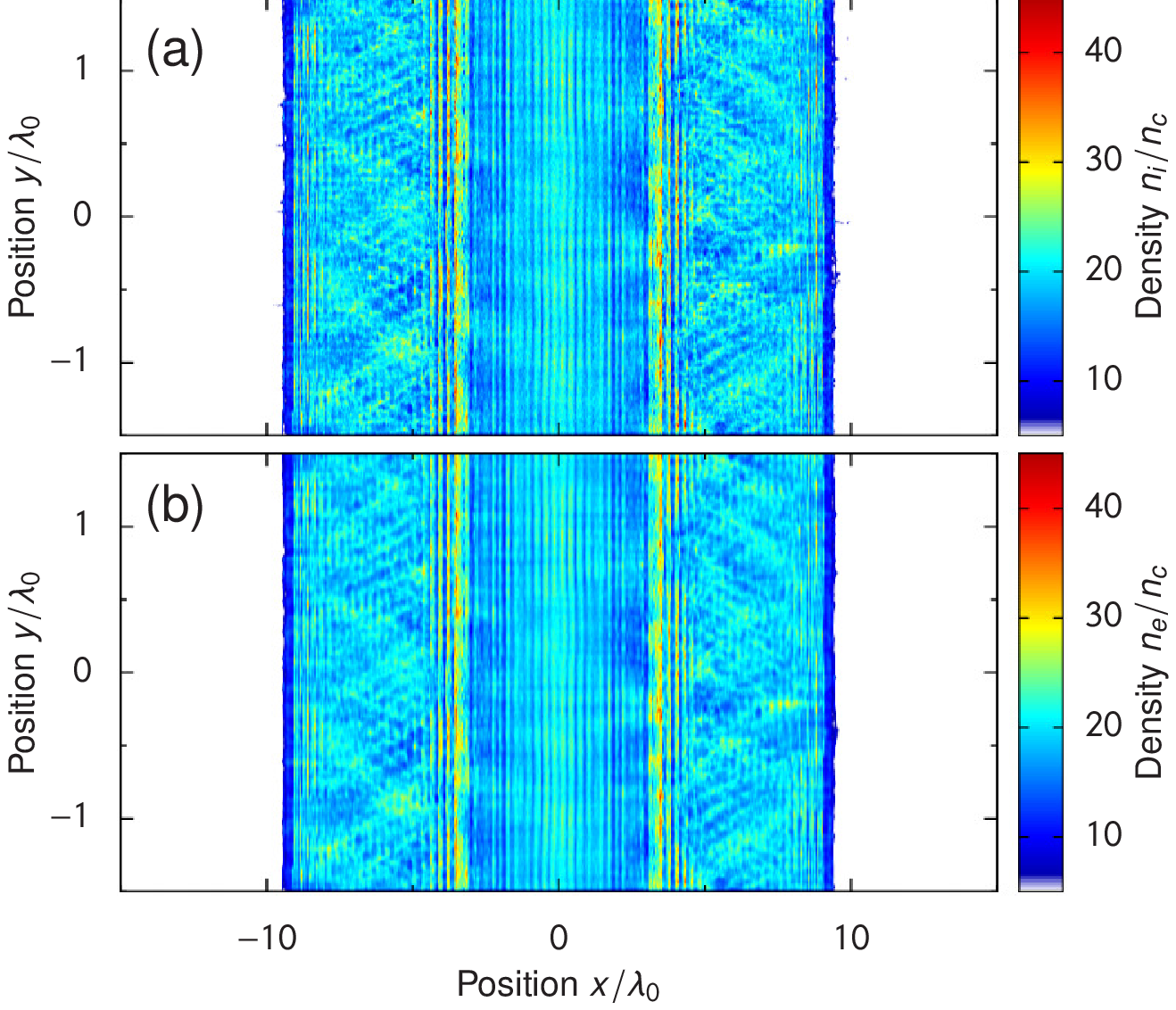}%
\caption{
Spatial distribution of the density $n$ for (a) ions and (b)
electrons in the 2D run. 
The initial parameters and snapshot timing are the same as those in
the 2D run shown by Fig~\ref{fig9}.
\label{fig9}}
\end{figure*}

The ion-heating mechanism by standing whistler waves is observed in
2D PIC simulations, which are performed to compare with the 1D
behavior.  
The initial parameters are the same as in the 1D
fiducial run except for the target thickness $L_x / \lambda_0 = 18.75$
and the pulse duration $\omega_0 \tau_0 = 35.3$.
The resolution in 2D runs is $\Delta x = c \Delta t = \lambda_0/300$
and the particle number per cell is 60.
The computational box size of the additional spatial direction $y$ is
considered from $L_y / \lambda_0 = 0.1$ to 3.
In the $y$ direction, the wave injection from the boundaries is uniform, and the periodic boundary condition is adopted.

Dependence of the energy conversion rate on the domain size in the
$y$ direction is shown in Fig.~\ref{fig6}.
Obviously the ion evolution is independent of $L_y / \lambda_0$.
The electron energy increases when $L_y / \lambda_0$ becomes comparable
to the whistler wavelength, which is $\lambda_w / \lambda_0 \sim 0.5$
for the fiducial parameters. 
However, it seems to be saturated if $L_y \gg \lambda_w$.
The same trend is recognized in the comparison of the energy spectra
between 1D and 2D simulations (Fig.~\ref{fig7}). 
All the spectra are well fitted by the thermal distribution of a
single temperature. 
The ion temperature estimated by the Maxwellian fitting is 22 keV for
the 1D run and 18 keV for 2D so that the ion spectra are unaffected
by 2D effects. 
The electron spectrum in 2D exhibits higher energy than that in 1D.
The electron temperature is 7.2 and 23 keV for 1D and 2D runs,
respectively. 

The ion energy is enhanced only at the central part of the target,
$| x / \lambda_0 | \lesssim 5$, where a standing whistler wave is
formed. 
Figure~\ref{fig8} shows the spatial distributions of the energy
density $u$ for ions and electrons in a 2D simulation with $L_y /
\lambda_0 = 3$.
It looks almost 1D-like distribution, and thus the ion evolution is
quite similar to the 1D case.
On the other hand, the electron energy is nearly uniform after the
passage of the injected whistler waves.
Because of the transverse propagation of the electron plasma waves,
the electrons absorb a larger amount of energy from the waves compared with the corresponding 1D result.
The evidence of the transverse plasma waves is observed in the spatial
distribution of density fluctuations during the standing whistler wave
heating (Fig.~\ref{fig9}).   
The initial parameters and the snapshot timing are the same as in
Fig.~\ref{fig8}. 
The standing whistler wave causes vertical stripes in the middle part.
Small-scale fluctuations of the order of the whistler wavelength
are seen in the transverse direction.



%





\end{document}